\documentclass[aps,preprint,preprintnumbers,amsmath,amssymb]{revtex4}
\usepackage{amsmath,mathrsfs,amsbsy,color,graphicx,bm,amsthm,amsfonts}
\usepackage{units}
\usepackage{bbm}
\usepackage{times}
\usepackage{dcolumn}
\usepackage{mathrsfs}
\usepackage{amsmath,amssymb,epsfig}
\usepackage{hyperref}

\DeclareMathOperator{\diag}{diag}

\begin{document}

\title{Gravity-enhanced quantum spatial target detection}
\author{Qianqian Liu$^{1}$, Cuihong Wen$^{1}$\footnote{Email: cuihongwen@hunnu.edu.cn}, Zehua Tian$^{2}$,  Jiliang Jing$^{1}$\footnote{Email: jljing@hunnu.edu.cn}, Jieci Wang$^{1}$\footnote{Email: jcwang@hunnu.edu.cn}}
\affiliation{$^1$  Department of Physics and Synergetic Innovation Center for Quantum Effects,\\Key Laboratory of Low-Dimensional Quantum Structures and Quantum Control of Ministry of Education, \\
 Hunan Normal University, Changsha 410081, China\\
 $^2$Hefei National Laboratory for Physical Sciences at the Microscale and Department of Modern Physics, Synergetic Innovation Center of Quantum Information and Quantum Physics,
University of Science and Technology of China, Hefei 230026, China}

\begin{abstract}

Quantum illumination can utilize entangled light to detect the low-reflectivity target that is hidden in a bright thermal background. This technique is applied to the detection of an object in the curved spacetime of the Earth, in order to explore how  the curvature of spacetime affects  quantum illumination. It is found that the spatial quantum illumination with entangled state transmitter outperforms that with coherent-state transmitter in the near-Earth curved  spacetime.
Moreover, either the quantum illumination system or the coherent-state system is employed, and gravity can enhance the spacetime target detection by reducing the thermal signal at the receiver.
 Besides, our model in principle can be applied to microwave quantum illumination and thus provides, to some degree, a theoretical foundation for the upcoming spatial quantum radar technologies.

\end{abstract}
\vspace*{0.5cm}

\maketitle
\section{Introduction}\label{section1}
Quantum illumination (QI) \cite{Lloyd} is a quantum-optical sensing technology that utilizes quantum entanglement \cite{Tan,Zhang1,Zhang2,Choi,Lopaeva,Ragy} to improve the detection of low-reflectivity targets under a bright thermal background.
Usually, two entangled optical modes are prepared in this scenario. One optical beam, as a signal  beam, is utilized to irradiate the target region, and the other, as an ancilla, is retained.
Whether or not the target exists depends on the returned mode from the target region and the retained idler \cite{Lloyd}.
 It was  experimentally demonstrated that the QI  outperforms the classical protocol with the enhancement
of quantum entanglement \cite{Zhang2,Z2,Z3}.
Quantum enhanced strategies can be applied to more occasions such as quantum
sensing  in  lossy and noisy environments \cite{Zhang1, Zhang2} and short-range radar technology \cite{Choi,Ren,RK1}.
Recently, Barzanjeh \emph{et al.} \cite{Barzanjeh,Barzanjeh2} proposed
an approach for microwave quantum illumination (MQI) using digital receiver.
Thanks to the naturally occurring bright thermal background in the microwave regime,  most target detection radars are available in this regime.
The quantum advantage of MQI in quantum radar technology is studied.

The preparation of quantum states and  the implementation of quantum information tasks are inevitably affected by relativistic effects since the gravitational interaction is non-maskable.
 In this perspective, it is of great significance to study the gravitational effect of the Earth on quantum systems,  such as the role of gravity in the quantum laboratory \cite{earthrev}  and the space quantum information processing tasks \cite{earthqs1,earthqs2,earthqs3, earthqs4,earthqs5,earthqs6,earthqs7}.
It is believed that studies of quantum information in relativistic framework can  provide new insights into some key issues in quantum mechanics and relativity, such as nonlocality, causality and the information paradox of black hole. 
To this end, a quantum information framework had been introduced for the description of photon propagation under the Earth's gravity \cite{earthqs1}.
The method was employed to study quantum communication \cite{earthqs1,earthqs4},  quantum metrology \cite{earthqs2,earthqs5,earthqs6}, quantum correlation \cite{earthqs3, earthqs7,QC}  and quantum clock synchronization \cite{wangqcs} under the influence of the Earth's spacetime curvature. The transmission of signal  beam involves transmitter-target range for near-Earth space target detection. Therefore, the spacetime effect of the Earth should be taken into serious consideration \cite{Rideout,Zych}.

This paper proposes a toy model of spatial QI  that describes the detection of a target located at different heights in a static gravitational potential. It demonstrates how the Earth's  gravity affects the QI with Gaussian states and gives a comparison between its performance and its flat spacetime counterpart.
It is assumed that one component of an entangled light wave-packet is sent from the Earth to a spatial target region where a target may exist, while the other component is retained to perform a joint measurement.
 The wave-packet overlap and the overall transmissivity are deformed by the Earth's spacetime curvature during the propagation process. Then the joint quantum measurement is performed on the returned signal and the ancilla beam, which presents how  the spacetime curvature dictates the spatial quantum target detection. It is found that the spacetime curvature of the Earth can weaken the contribution of thermal-noise to the returned signal, which helps the gravitational effects of the Earth enhance the probability of successful detection.

This paper is structured as follows. Section \ref{section2} describes the propagation of the photons under the background of the Earth.
 Section \ref{section3} is devoted to analyzing the advantages of the Earth's  spacetime curvature to QI strategy and coherent-state strategy.
 Finally, the conclusions are drawn in Section \ref{section4}.

\section{Light wave-packets propagation in Earth spacetime}\label{section2}


The model of light wave-packets propagation in the Earth spacetime
has been introduced in detail in Refs. \cite{earthqs1,earthqs2}.
 The Earth spacetime can be approximately described by the Kerr metric  \cite{Visser}. For simplicity, our work will be constrained to the equatorial plane $\theta=\frac{\pi}{2}$. The reduced metric in
the Boyer-Lindquist line element reads \cite{Visser}
\begin{align}\label{metric}
ds^2=&\, -\Big(1-\frac{2M}{r} \Big)dt^2+\frac{1}{1-\frac{2M}{r}+\frac{a^2}{r^2}}dr^2+\Big(r^2+a^2+\frac{2Ma^2}{r}\Big) d\phi^2 - \frac{4Ma}{r} dt \, d\phi,
\end{align}
where we will consider the rotating plant to be the Earth, with radius $r_{A}$, mass $M_{A}$ and angular momentum $J$ respectively, with the Kerr parameter $a=\frac{J}{M_{A}}$.
$\hbar=G=c=1$ being set throughout this paper.

A photon can be properly modeled by a wave-packet of electromagnetic fields with a distribution $F^{(K)}_{\Omega_{K,0}}$,
where $K=A,B$ labels the users at different distances from the Earth. Here $\Omega_K$ is the physical frequency as measured in the corresponding transmitter and the peak frequency $\Omega_{K,0}$ \cite{Downes,Leonhardt}.
The annihilation operator for the photon, from different locations of observers, takes the form
\begin{eqnarray}\label{operators}
\hat{a}_{\Omega_{K,0}}(t_K)=\int_0^{+\infty}d\Omega_K\, e^{-i\Omega_K t_K} F^{(K)}
_{\Omega_{K,0}}(\Omega_K)\,\hat{a}_{\Omega_K}.
\end{eqnarray}
If the frequency distribution $F^{(K)}
_{\Omega_{K,0}}(\Omega_K)$ is normalized, i.e. $\int_{\omega>0} d\omega |F^{(K)}
_{\Omega_{K,0}}(\Omega_K)|^2=1$, such a distribution naturally satisfies a pulse where the optical field is described as a local propagation model in spacetime.
The annihilation operator $\hat{a}_{\Omega_{K}}$ and creation operator $\hat{a}^{\dag}_{\Omega_{K}}$ must satisfy the following canonical commutation relation
\begin{eqnarray}
[\hat{a}_{\Omega_K},\hat{a}^{\dagger}_{\Omega_K'}]=\delta(\Omega_K-\Omega'_K).
\end{eqnarray}
Then we consider the effect of spacetime on the propagation of the photons.
If one observer Alice at time $\tau_A$ and location $r_A$, sends a wave-packet $F^{(A)}_{\Omega_{A,0}}$ to the other observer Bob at location $r_B$. Bob will receive at the time $\tau_{B}=\Delta\tau+\sqrt{f(r_{B})/f(r_{A})}\tau_{A}$,
where $\Delta\tau$ represents the propagation time of the light, and  $f(r_{A(B)})$ is the gravitational frequency shifting factor at different heights.
 The modified wave-packet received by Bob is denoted by $F^{(B)}_{\Omega_{B,0}}$.
Notably, the operators in Eq. (\ref{operators}) can be used to describe the same optical mode at two different locations before and after the propagation.

As shown in Refs. \cite{earthqs1,earthqs2,earthqs3,earthqs4,earthqs5,earthqs6,DE,DE2},
we can utilize the relation between the annihilation operator $\hat{a}_{\Omega_A}$ and $\hat{a}_{\Omega_B}$ to obtain the relation between
the frequency distributions $F^{(K)}_{\Omega_{K,0}}$ in different reference frames (or before and after propagation)
\begin{eqnarray}
F^{(B)}_{\Omega_{B,0}}(\Omega_B)=\sqrt[4]{\frac{f(r_B)}{f(r_A)}}
F^{(A)}_{\Omega_{A,0}}\left(\sqrt{\frac{f(r_B)}{f(r_A)}}
\Omega_B\right).\label{wave:packet:relation}
\end{eqnarray}
Obviously, the wave-packet received by the observer Bob has a peak frequency and a shape that are different from the wave-packet prepared by the observer Alice.
 In Refs. \cite{earthqs1,earthqs2},  the authors had demonstrated that
 the gravitational effects affect quantum communication in space and quantum metrology of the Schwarzschild radius (mass of the Earth).

In fact, the  propagation of the photons between two different locations in curved spacetime is similar to a beam splitter operation performing on propagating photons. To be specific, it's possible for the mode $\bar{a}^{\prime}$
to be decomposed in terms of the mode $\hat{a}^{\prime}$ and the orthogonal mode $\hat{a}_{\bot}^{\prime}$ \cite{earthqs1,earthqs2, Rohde,earthqs7}
\begin{eqnarray}
\bar{a}^{\prime}=\Theta
\hat{a}^{\prime}+\sqrt{1-\Theta^2}\hat{a}_{\bot}^{\prime},\label{mode:decomposition}
\end{eqnarray}
where $\Theta$ is the wave-packet overlap  between the distributions
$F^{(B)}_{\Omega_{B,0}}(\Omega_B)$ and
$F^{(A)}_{\Omega_{A,0}}(\Omega_B)$,
\begin{eqnarray}
\Theta=\int_{0}^{+\infty}d\Omega_B\,F^{(B)\star}_{\Omega_{B,0}}(\Omega_B)F^{(A)}_{\Omega_{A,0}}(\Omega_B).\label{single:photon:fidelity}
\end{eqnarray}
The quality of the channel can be quantified by fidelity $\mathcal{F}=|\Theta|^{2}$. If the channel is perfect, $\Theta=1$.
The transformation in Eq. (\ref{mode:decomposition}) describes a lossy channel
which reflects photon loss  probability $1-\Theta^2$.
It's assumed that the wave-packet $F_{\Omega_0}(\Omega)$ is Gaussian distribution that satisfies
\begin{eqnarray}
F_{\Omega_0}(\Omega)=\frac{1}{\sqrt[4]{2\pi\sigma^2}}e^{-\frac{(\Omega-\Omega_0)^2}{4\sigma^2}}\label{Bobpacket},
\end{eqnarray}
with the wave-packet width $\sigma$.

The integral in Eq.  \eqref{single:photon:fidelity} should perform over strictly positive frequencies.
Nonetheless, since the peak frequency $\Omega_{0}$ is typically much larger than the wave-packet width $\sigma$, negative frequencies can be involved without affecting the overlap $\Theta$. Based on the
combination of Eq. \eqref{single:photon:fidelity} and \eqref{Bobpacket}, one could
obtain
\begin{eqnarray} \label{theta}
\Theta_{1(2)}=\sqrt{\frac{2(1\pm\delta)}{1+(1\pm\delta)^2}}e^{-\frac{\delta^2\Omega_{B,0}^2}{4(1+(1\pm\delta)^2)\sigma^2}}\label{final:result},
\end{eqnarray}
where the signs $\pm$ occur for the upwards (i.e., $r_{B}>r_{A})$ and downwards (i.e., $r_{B}<r_{A})$ processes. The new parameter introduced to quantify the
shifting is
\begin{equation}\label{delta}
\delta=\sqrt[4]{\frac{f(r_A)}{f(r_B)}}-1=\sqrt{\frac{\Omega_{B}}{\Omega_{A}}}-1.
\end{equation}
In the equatorial plane of the Kerr spacetime, the frequency ratio of the wave-packet before sending and receiving is \cite{earthqs4}
\begin{equation}\label{aw}
\frac{\Omega_{B}}{\Omega_{A}}=\frac{1+\epsilon
\frac{a}{r_B}\sqrt{\frac{M_{A}}{r_B}}}{C\sqrt{1-3\frac{M_{A}}{r_B}+
2\epsilon\frac{a}{r_B}\sqrt{\frac{M_{A}}{r_B}}}},
\end{equation}
where
$C=[1-\frac{2M_{A}}{r_A}(1+2a
{\omega})+\big(r^2_A+a^2-\frac{2M_{A}a^2}{r_A}\big){\omega}^2]^{-\frac{1}{2}}$ is the normalization constant
(with the Earth's equatorial angular velocity $\omega$), and
$\epsilon=\pm1$ stands for the direction of the orbits.
To obtain the expression of the frequency shift between Alice and Bob affected by gravity, we expand Eq. (\ref{aw}) and keep the second-order term of the parameter $r_{A}\omega$.
The perturbative result is independent of the states of the Earth and the  satellite (i.e., whether they are
co-rotating or not)
\begin{eqnarray}\label{bw}
\delta&=&\delta_{Sch}+\delta_{rot}+\delta_h\\
\nonumber&=&\frac{1}{8}\frac{r_S}{r_A}\big(\frac{r_A-2R}{r_A+R} \big)-\frac{(r_A\omega)^2}{4}-\frac{(r_A\omega)^2}{4}\big(\frac{3}{4}\frac{r_S}{r_A}-\frac{2r_Sa}{\omega r_A^3}\big),
\end{eqnarray}
where the parameter $R=r_B-r_A$
is the height difference between Bob and Alice. In Eq. \eqref{bw},
$\delta_{Sch}$, $\delta_{rot}$ and $\delta_h$
denote the first order Schwarzschild term, the rotation term and
the higher-order correction term,  respectively. And $r_S=2M_{A}$ is the
Schwarzschild radius of the Earth. If the satellite is located at the height $R\simeq\frac{r_{A}}{2}$, $\delta=0$, $\Theta=1$ can be drawn,
which indicates that the photons received at this height possess no frequency shift.

Considering that the radius of the Earth is about $r_{A}=6371$ km, and its Schwarzschild radius is about $r_S=9$ mm, we note that in Eq. \eqref{theta} two different scenarios occur \cite{earthqs1}.
(i) If $\frac{\delta\Omega_{B,0}}{\sigma}\leq\delta\ll1$, (for example, $\delta\leq10^{-14}$ is attained
 if the target ranges around 100m),  then the wave-packet overlap is $\Theta\sim1-\mathcal{O}(\delta^{2})\sim1$. In this case, the gravitational effects are independent of the peak frequency and on the bandwidth of the distribution and can be ignored.
 (ii) The other case is $\delta\ll(\frac{\delta\Omega_{B,0}}{\sigma})^{2}\ll1$,
  which occurs for typical communication where $\Omega_{B,0}=700 \textrm{THz}$ and Gaussian bandwidth ${\sigma}=1\textrm{MHz}$ (the similar
peak frequencies and bandwidths have been reported recently by trapped ion experiments \cite{700THZ}).
  In this case, the wave-packet overlap parameter is $1-\frac{\delta^{2}\Omega_{B,0}^{2}}{8\sigma^{2}}$.
  The  effect of gravity  is much larger than the first scenario, and it should be taken into account in the propagation of photons.
Accordingly, the final state and the wave-packet overlap $\Theta$ are related to the range $R$ of the target.
We assume that the signal photon propagation has additional bright light interference \cite{story}, which accumulates the advantage of the ideal QI radar in the optical band.

\section{Quantum illumination in curved spacetime}\label{section3}

\begin{figure}[ht]
\centering
\includegraphics[width=0.58\textwidth]{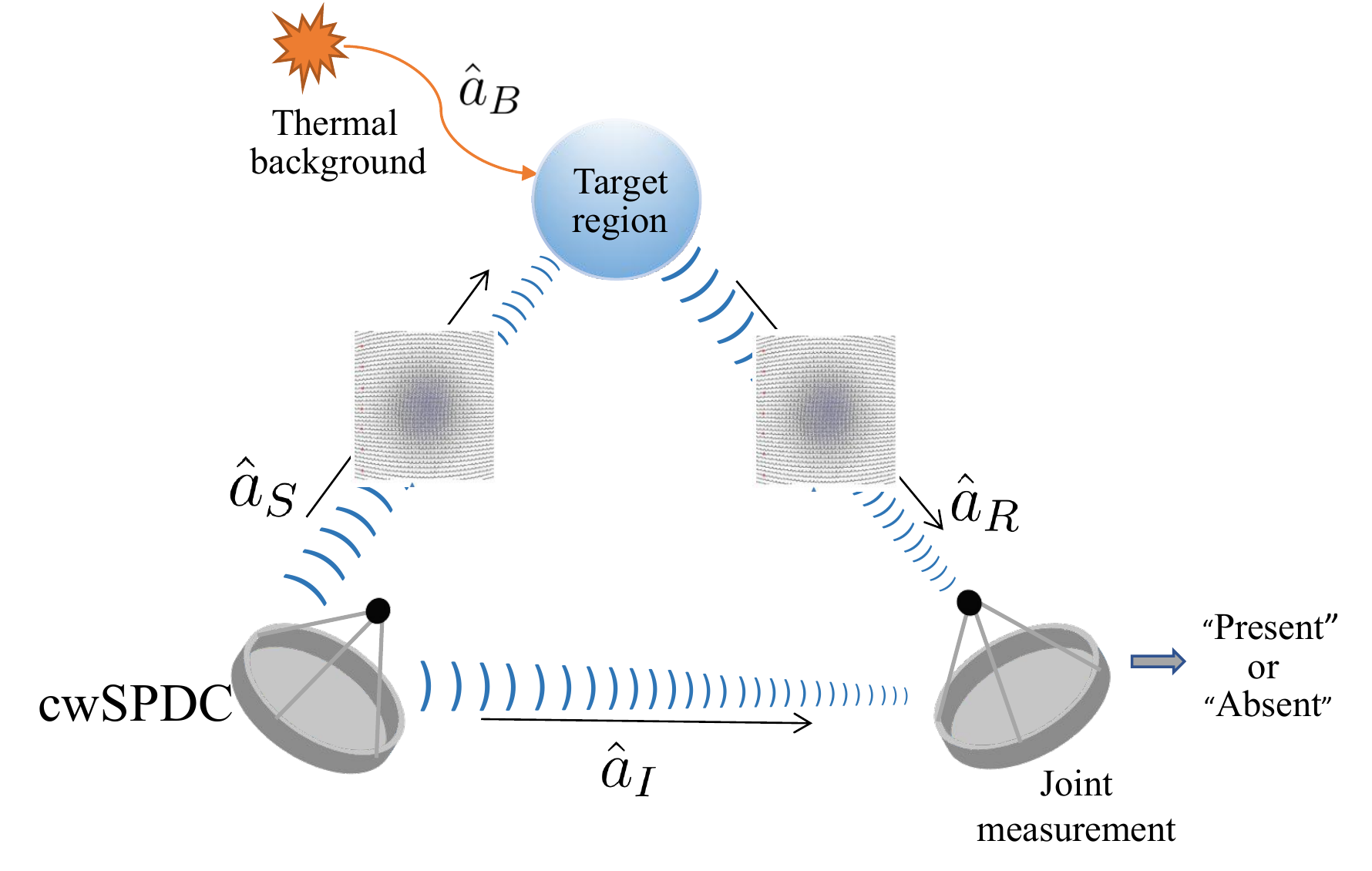}
\caption{(Color online) QI under the curved spacetime background  of the Earth. A photon $\hat{a}_{S}$ of the entangled light wave-packet is sent from the Earth to a target region, and the mode $\hat{a}_{R}$  returned  from the target region is used to decide whether the target is present or absent.}\label{fig1}
\end{figure}

 Fig. (\ref{fig1}) shows the diagram of our proposed strategy for QI under the Earth's gravitational effect. Unlike the original QI model proposed in  \cite{Tan}, the propagation of photons in curved spacetime is considered here.
 Continuous-wave (cw) spontaneous parametric down-conversion (SPDC)  \cite{Wong} produces a pair of entangled photons.
One of the entangled photons forming the signal beam $\hat{a}_{S}$ irradiates a spatial target region. The other photon is retained as the  idler mode  $\hat{a}_{I}$  to perform a joint quantum measurement with the returned mode.
 The sources for noise photons in the environment are countless and varied, including the sunlight, atmospheric counter radiation, urban illumination, etc. For simplicity, the noise background  in  our model  is assumed to be a thermal-noise bath.
Due to the curvature of spacetime, the propagation of photon is affected via changing their frequency distribution in center, shape and bandwidth. Because of the change of the above mode, the final joint measurement is different from that in the flat spacetime.

\newpage

The interaction of the transmitter with the target may be modeled as a beam splitter with transmissivity $\kappa$ \cite{Lloyd,Tan,RK1}.
In the hypothesis of the target  presents $(H_1)$, transmissivity $\kappa$
includes the impact of the path loss and non-directional antennas.
Although the probability of a signal photon returned by a target is quite low, the occurrence of such an event will play a crucial role in the subsequent joint measurement.
In the hypothesis of the target absent $(H_0)$, a beam splitter scheme with $\kappa=0$ is obtained.
In this case,
all the signal modes filter through the target region without any reflection, and the returned mode is completely replaced by the optical mode of thermal-noise. Then we obtain a very strong background contribution at the receiver, which is independent of the target's presence or absence.
If both hypotheses are equally-likely, Helstrom \cite{Helstrom} had given a minimum error-probability decision rule  for the QI strategy.

  Through measuring $\rho^{(1)}(\kappa\neq0)-\rho^{(0)}(\kappa=0)$, it is argued that the target is present if the result is a non-negative number; otherwise, the target is absent.
  Then the error-probability of this
optimum quantum receiver for the quantum target detection transmitter is
\begin{eqnarray}\label{error-probability}
\mathrm{Pr}(e)=\frac{1}{2}\big(1-\sum_{n}\gamma_{n}^{(+)}),
\end{eqnarray}
where $\gamma_{n}^{(+)}$ denotes the non-negative eigenvalues of $\hat{\rho}^{(1)}-\hat{\rho}^{(0)}$.
To better discriminate the quantum states under each of the two hypotheses,  the study adopted  the multicopy states $\hat{\rho}^{(1)\otimes\,M}$ and $\hat{\rho}^{(0)\otimes\,M}$ and utilized the optimal joint quantum
measurement \cite{Tan}. Then the mean error-probability is found to be
\begin{eqnarray}\label{error-probability2}
\nonumber \mathrm{Pr}(e)=\frac{1}{2}\big(1-\frac{1}{2}\parallel\hat{\rho}^{(1)\bigotimes M}- \hat{\rho}^{(0)\bigotimes M} \parallel),
\end{eqnarray}
where $\emph{M}$ is the number of entangled light sources used for QI detection, and  $\|A\|=\textmd{Tr}\sqrt{A^{\dag} A}$ represents the trace norm of quantum state $A$.
Calculating the trace norm in Eq. (\ref{error-probability}) is a daunting task, yet the quantum Chernoff bound \cite{Audenaert, Nussbaum, Calsamiglia} sets a limit on the error-probability
\begin{eqnarray}\label{CF-bound}
\mathrm{Pr}(e)\leq\frac{1}{2}e^{-M\epsilon}\equiv\frac{1}{2}\{\min_{0\leq\,s\leq1}
\mathrm{Tr}[(\rho^{(0)})^s(\rho^{(1)})^{(1-s)}]\}^M.
\end{eqnarray}
The bound is exponentially tight, i.e., the error-probability exponent $-\ln[\mathrm{Pr}(e)]/M$ converges to $\epsilon$ as
$M\rightarrow\infty$. In the potentially weaker case ( $s=1/2$), one can obtain the Bhattacharyya bound \cite{Kailath, Pirandola}
\begin{eqnarray}\label{inequality1}
\mathrm{Pr}(e)\leq\frac{1}{2}\{\mathrm{Tr}[(\hat{\rho}^{(0)})^{1/2}(\hat{\rho}^{(1)})^{1/2}]\}^M.
\end{eqnarray}
In order to obtain more amenable analytic results, we combine the lower bound of the error-probability
\begin{eqnarray}\label{inequality1}
\mathrm{Pr}(e)\geq\frac{1}{2}(1-\sqrt{1-\{\mathrm{Tr}[(\hat{\rho}^{(0)})^{1/2}(\hat{\rho}^{(1)})^{1/2}]\}^{2M}}),
\end{eqnarray}
in general, which is not exponentially tight.

In the following, we will investigate the target detection with the QI transmitter and  coherent-state transmitter.
 It is feasible to evaluate the Chernoff  or Bhattacharyya bounds for Gaussian quantum states with the help of Ref. \cite{Pirandola}.
For the QI transmitters \cite{Tan}, an entangled signal and idler mode pair are obtained
from cw SPDC. This mode pair with
annihilation operators $\hat{a}_S$ and $\hat{a}_I$ is in the entangled state with number-ket representation
\begin{eqnarray}\label{IS-state}
|\phi\rangle_{\mathrm{IS}}=\sum^\infty_{n=0}\sqrt{\frac{N^n_S}{(N_S+1)^{n+1}}}|n\rangle_\mathrm{I}|n\rangle_\mathrm{S},
\end{eqnarray}
where $N_S$ is the average photon number per mode. We work on the phase space, where quantum states can be described by quadrature operators $\hat{R}={(\hat{x}_{1},\hat{p}_{1},\hat{x}_{2},\hat{p}_{2},...,\hat{x}_{n},\hat{p}_{n})^{T}}$.  Such  operators  are related to the annihilation $\hat{a}_{i}$ and creation $\hat{a}^{\dag}_{i}$ operators by the relations $\hat{x}_{i}=\frac{(\hat{a}_{i}+\hat{a}^{\dag}_{i})}{\sqrt{2}}$ and $\hat{p}_{i}=\frac{(\hat{a}_{i}-\hat{a}^{\dag}_{i})}{\sqrt{2i}}$,  for each mode $i$.
The quadrature operator $\hat{R}$ satisfies the commutation relation: $[\hat{R}_{k},\hat{R}_{l}]=i\Omega_{k,l}$, with $\Omega=\bigoplus^{n+m}_{1} {{\ 0\ \ 1} \choose{-1 \ 0}}$ being symplectic matrix.
The properties of a Gaussian state are completely specified by the first moment and the second moment.
 In the phase space representation,
$|\phi\rangle_{\mathrm{IS}}$ is a zero-mean Gaussian state whose corresponding
covariance matrix \cite{Weedbrook,Tan} is
\begin{eqnarray}\label{IS-CM}
\Lambda_\mathrm{IS}=\left[
\begin{array}{cccc}
  C & 0 & S & 0 \\
  0 & C & 0 & -S \\
  S & 0 & C & 0 \\
  0 & -S & 0 & C
\end{array}
\right],
\end{eqnarray}
 where $C=2N_S+1$ and $S=2\sqrt{N_S(N_S+1)}$.

For the QI strategy,
the entangled signal $\hat{a}_S$ and $\hat{a}_I$  are prepared in the cw SPDC.
 In the present proposal, an entangled state transmitter illuminates a region containing a bright thermal-noise bath, in which a low-reflectivity target might be embedded. The overall  process is carried out under the influence of the Earth.
 If the target is not detected, the returned signal will be composed of the thermal signal $\hat{a}_{B}$.
 If a low-reflectivity target is detected, the signal photon will be affected by the thermal signal, so the returned signal is a combination of $\hat{a}_{S}$ and $\hat{a}_{B}$.  In either case, the space contains a bright thermal-noise bath.
We take a joint measurement of the mode $\hat{a}_{I}$ and $\hat{a}_{R}$ returned from the target region.
Under hypothesis $H_0$ (target absent), the mode returned from the target region will be
 \begin{eqnarray}
 \hat{a}_{R_0}={\Theta} \hat{a}_B+\sqrt{1-\Theta^{2}}\hat{a}_{B\perp},
  \end{eqnarray}
  where $\hat{a}_B$ is the annihilation operator of the thermal-noise mode.
  Under hypothesis $H_1$ (target present), the respective effects of the gravitational red-shift and the gravitational blue-shift on the illuminating signal $\hat{a}_S$ are canceled with each other. However, the gravity of the Earth still has impact on the signal coming from the thermal background that is transmitted from the target to the receiver. Then the annihilation operator for the returned mode will be
  \begin{eqnarray}
 \hat{a}_{R_1}=\sqrt{\kappa}\hat{a}_S+\sqrt{1-\kappa}({\Theta} \hat{a}_B+\sqrt{1-\Theta^{2}}\hat{a}_{B\perp}).
 \end{eqnarray}

  In the present model, the overall transmissivity $\kappa$ is the ratio between the  received power $P_{R}$ and the transmitted power $P_{T}$ \cite{RK1},
  \begin{equation}
\kappa=\frac{P_{R}}{P_{T}}=\frac{G F^{4} A_{R} \sigma'}{(4 \pi)^{2} R^{4}}.
\end{equation}
where $\sigma'$ and $G$ are the cross-section of the target and gain of the transmit antennas, and $R$ is the transmitter target range. The other two factors, $A_{R}$ and $F$ are
the effective areas of the radar receive antenna and the form factor, respectively. Here the factor $\frac{1}{(4 \pi)^{2} R^{4}}$ represents the loss of the pulse signal during the whole propagation process.
In addition, for a short range QI, the beam spreading does not involve too much loss, which is the major killing factor for any practical quantum target detection tasks.
In fact, the QI proposal  can be extended to variable distances by sending entangled pulses at different carrier frequencies and measuring their reflection at different roundtrip times.

 In the short range scenario, we can assume $F=1$ (no free space loss)  and an ideal pencil beam, so that its solid angle $\delta'$ is exactly subtended by the target's cross section $\sigma'$ \cite{RK1}.
This means that gain factor can be technically given by
 \begin{equation}\nonumber
G=\frac{4\pi}{ \delta'}=\frac{4\pi R^{2}}{\sigma'}.
\end{equation}
Then the overall transmissivity is found to be
\begin{equation}
\kappa=\frac{ A_{R} }{(4 \pi R)^{2}}.
\end{equation}
We can obtain a one-to-one correspondence between transmissivity $\kappa$ and range $R$ by fixing the receive antenna collecting area $A_{R}=0.1\textrm{m}^{2}$.

 It is noticed that under both hypotheses, the joint state of the $\hat{a}_{R}$ and $\hat{a}_{I}$ modes is a zero-mean mixed Gaussian state. The corresponding  covariance matrices under $H_0$ and $H_1$ are
\begin{eqnarray}\label{H0}
\Lambda^{(0)}_{\mathrm{IR}}=\left[
\begin{array}{cccc}
  C & 0 & 0& 0 \\
  0 & C & 0 & 0 \\
  0 & 0 & 1-\Theta^{2}+\Theta^{2}B & 0 \\
  0 & 0& 0 & 1-\Theta^{2}+\Theta^{2}B
\end{array}
\right],
\end{eqnarray}
and
\begin{eqnarray}\label{H1}
\Lambda^{(1)}_{\mathrm{IR}}=\left[
\begin{array}{cccc}
  C & 0 & S' & 0 \\
  0 & C & 0 & -S' \\
  S' & 0 & 1-\Theta^{2}+A & 0 \\
  0 & -S' & 0 & 1-\Theta^{2}+A
\end{array}
\right],
\end{eqnarray}
respectively, where $S'=\sqrt{\kappa}S$, $A=2\kappa\,N_S+\Theta^{2}B$ and $B=2N_B+1$. To obtain the error-probability, it is necessary to diagonalize the Gaussian states Eqs. (\ref{H0}) and (\ref{H1}) with appropriate symplectic matrices. For $\Lambda^{(0)}_{\mathrm{IR}}$, its four dimension symplectic matrix is an identity matrix, and the associated symplectic eigenvalues are $\nu_1=C$ and $\nu_2=1-\Theta^{2}+\Theta^{2}B$.
One can compute the
symplectic matrix \cite{Weedbrook} to diagonalize $\Lambda^{(1)}_{\mathrm{IR}}$, which is given by
\begin{eqnarray}\label{s-matrix1}
\mathcal{M}=\left(
\begin{array}{cc}
  \omega_+\mathbf{1} & \omega_-\mathbf{Z} \\
  \omega_-\mathbf{Z} & \omega_+\mathbf{1}
\end{array}
\right),
\end{eqnarray}
with
$\mathbf{1}=\diag\{1,1\}$, $\mathbf{Z}=\diag\{1,-1\}$, and
\begin{eqnarray}
\omega_\pm=\sqrt{\frac{C^\prime\pm\sqrt{C^{\prime2}-4\kappa\,S^2}}
{2\sqrt{C^{\prime2}-4\kappa\,S^2}}},
\end{eqnarray}
where $C^\prime=C+A+1-\Theta^{2}$ is assumed. In this case, the associated symplectic eigenvalues can be denoted by
\begin{eqnarray}
\nu_n=\frac{1}{2}\bigg[(-1)^n(2C-C^\prime)+\sqrt{C^{\prime^2}-4\kappa\,S^2}\bigg],
\end{eqnarray}
with $n=1, 2$.
These symplectic eigenvalues can be employed to attain an asymptotical expression for the QI transmitter's Bhattacharyya bound when $\kappa\ll1$, $N_{S}\ll1$, and  $N_{B}\gg1$:
\begin{eqnarray}
\mathrm{Pr}(e)_{\mathrm{QI}}\leq\frac{1}{2}\exp[-M\kappa\,N_S/\Theta^{2}N_B].
\end{eqnarray}

If we employ a single mode coherent-state to irradiate the target region, the covariance matrices can be obtained in a similar way.
 Under the assumptions that the target is either absent or present,  the forms of the covariance matrices of the returned mode are: $\Lambda^{(0)}=\diag\big\{ 1-\Theta^{2}+\Theta^{2}B, 1-\Theta^{2}+\Theta^{2}B\big\}$, $\Lambda^{(1)}=\diag\big\{ 1-\Theta^{2}+\Theta^{2}B, 1-\Theta^{2}+\Theta^{2}B\big\}$
and $\text{Tr}(\rho^{(0)}\hat{a}_R)=0$, $\text{Tr}(\rho^{(1)}\hat{a}_R)=\sqrt{\kappa\,N_S}$. In this case,
 the quantum Chernoff bound \cite{Pirandola}
 (which turns out to be the Bhattacharyya bound) is found to be
 \begin{eqnarray}\label{cb1}
\nonumber
\mathrm{Pr}(e)_{\mathrm{CS}_{1}}&\leq&\frac{1}{2}\exp[-M\kappa\,\Theta^{2}N_S(\sqrt{N_B+1/\Theta^{2}}-\sqrt{N_B})^2]
\\ \nonumber
&\approx&\frac{1}{2}\exp[-M\kappa\,N_S/4\Theta^{2}N_B],~~~\text{when $N_B\gg1$}.
\\
\end{eqnarray}
On the other hand, the lower bound of the error-probability of the coherent-state transmitter is also an important figure of merit \cite{Pirandola},
which is given by
\begin{widetext}
\begin{eqnarray}\label{cb2}
\nonumber
\mathrm{Pr}(e)_{\mathrm{CS}_{0}}&\geq&\frac{1}{2}\bigg(1-\sqrt{1-\exp[-2M\kappa\,\Theta^{2}N_S(\sqrt{N_B+1/\Theta^{2}}-\sqrt{N_B})^2]}\bigg)
\\
&\approx&\frac{1}{4}\exp[-M\kappa\,N_S/2\Theta^{2}N_B],~~~~~~\text{when $N_B\gg1$ and $M\kappa\,\Theta^{2}N_S/2N_B\gg1$}.
\end{eqnarray}
\end{widetext}

\begin{figure}[ht]
\centering
\includegraphics[height=2.5in,width=3.2in]{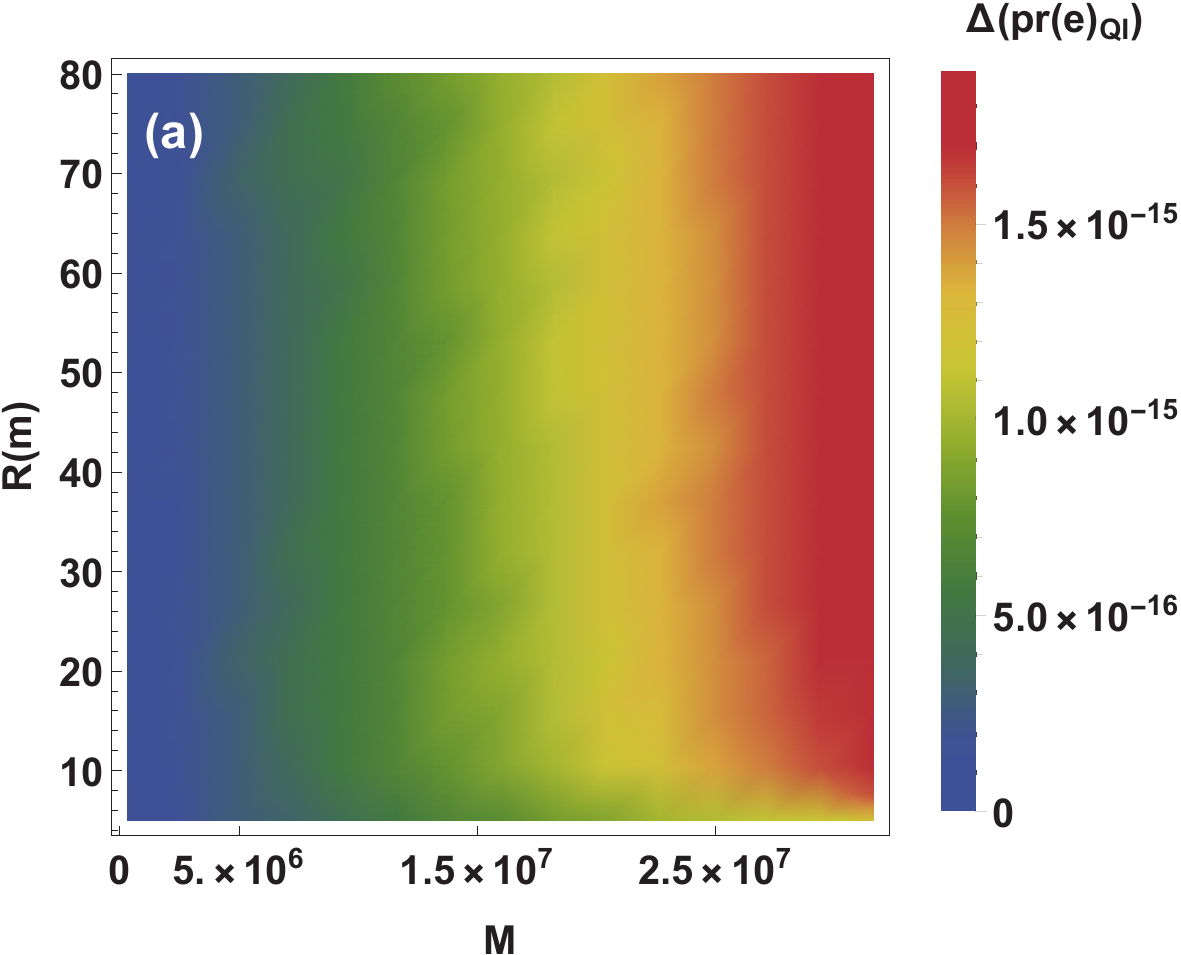}
\includegraphics[height=2.5in,width=3.2in]{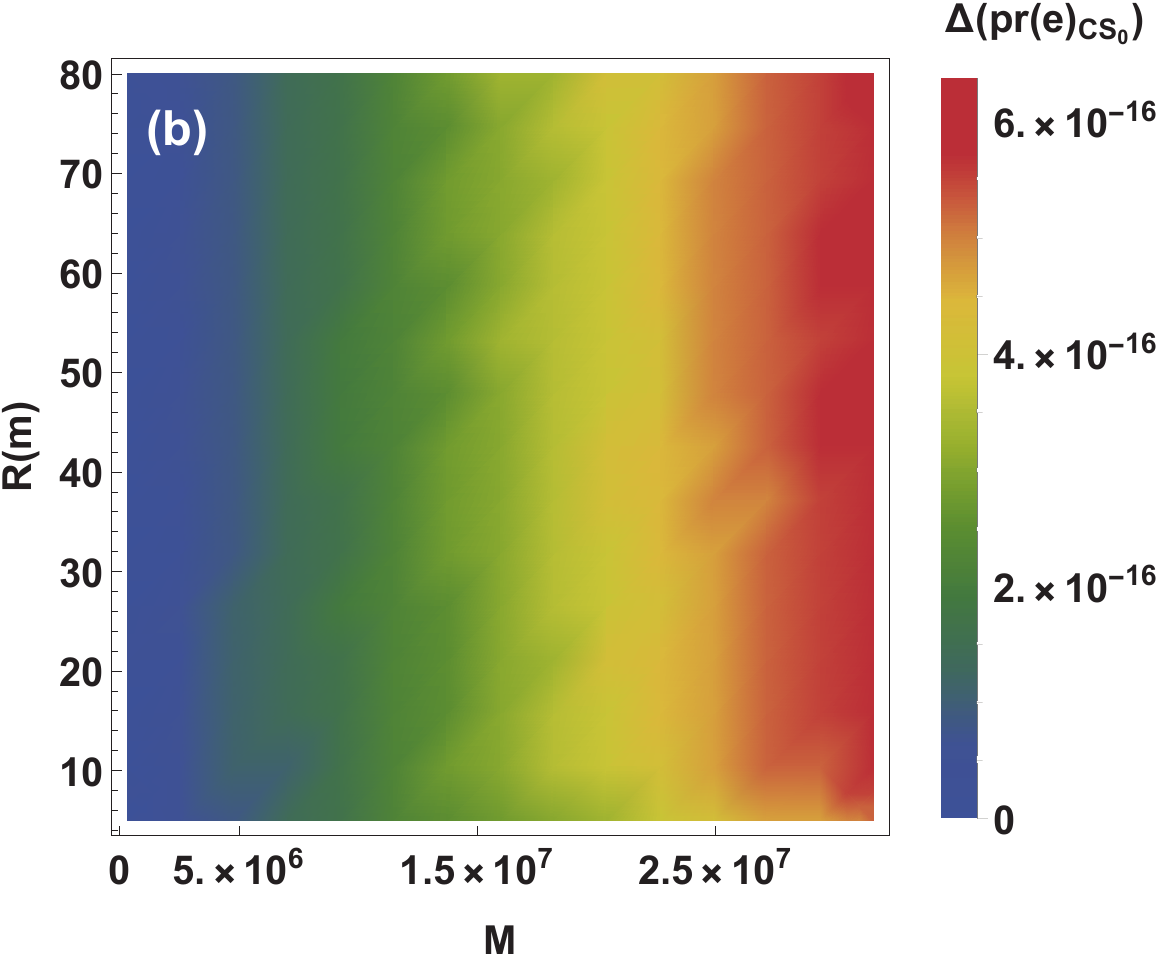}
\caption{(Color online) The advantage of the target detection error-probability in curved spacetime over flat spacetime of (a) the QI transmitter and (b) the coherent-state transmitter. Here the signal brightness $N_S=0.01$ and the environmental noise $N_B=20$. }
\end{figure}\label{fig22}

In Fig. 2(a), we plot the advantage of the target detection error-probability
$\Delta(\textrm{Pr(e)}_{\mathrm{QI}})$ of the QI transmitter in the near-Earth spacetime over the flat case. $\Delta(\textrm{Pr(e)}_{\mathrm{QI}})$  is a positive value in the figure, which indicates that the error probabilities in the near-Earth curved spacetime is lower than its flat counterpart. Furthermore, the advantage of QI systems
is positively associated with the copies of the transmitted modes $M$.
 The advantage  error-probability of the curved spacetime target detection for the coherent-state transmitter in Fig. 2(b). The Earth's gravity is able to reduce the detection error of the coherent-state transmitter.
When either the QI system or the coherent-state system is employed, the error-probability in the near-Earth curved spacetime case is always lower than that in the flat spacetime.
We also find that the advantage of the detection error-probability of the QI transmitter in curved spacetime is about $10^{-15}$ compared with that in the flat spacetime. These advantages outweigh the detecting precision of atom clocks \cite{atom}. Therefore, the Earth's gravity presents non-negligible effects on the QI proposal. This is not surprising since the time
dilation induced by the Earth's spacetime curvature is experimentally observed for a height change of 0.33m \cite{atom}. It is expected that the gravitational effects of the Earth would be more pronounced for MQI at longer distances.

We can see that  the Earth's gravity enhances the efficiency of the spatial target detection.
The result is nontrivial since gravitational effects usually destroy quantum correlation  such as  entanglement \cite{Ivette,Wang2, David3, Tian,JW}, and gravity reduces the fidelity of teleportation \cite{earthqs1} in relativistic quantum information.
Unlike the dynamic of a quantum system near the event horizon of a black hole \cite{Hawking}, the present model has nothing to do with the creation of particles.

We now proceed to give a physical interpretation of this phenomenon.
  The photon wave-packet between the reference $F^{(A)}_{\Omega_{A,0}}$ and the satellite observer $F^{(B)}_{\Omega_{B,0}}$ are the same in the case of flat spacetime ($R=0$).
 If the QI performs in the curved spacetime ($R>0$), the propagation of wave-packets will pass  through a lossy channel.
 Therefore,  the relationship between the  operator $\hat{a}_{S}$ for the signal photons and  the operator $\hat{a}_{B}$ for the thermal photons has been changed under the influence of the curved spacetime.
  The gravitational effects on the entangled signal $\hat{a}_{S}$ can  cancel each other out because of the gravitational red-shift (in upwards process) and the gravitational blue-shift (in downwards process).
However, for the thermal signal $\hat{a}_{B}$ returned from the background spacetime, it is changed by the gravitational effects of the Earth before it reaches the receiving end.
Comparing our final state with the flat space case \cite{Tan}, it is easy to find that  the original $B$ is replaced with $1-\Theta^{2}+\Theta^{2}B$ in our model.
It is shown that the contribution of the thermal signal returned in the curved spacetime is always less than that in flat.
That is to say, the spacetime curvature of the Earth can decrease the contribution of thermal-noise to the received signal, and the reduction of the thermal signal enables greater efficiency of the target detection.
  Due to the existence of wave-packets overlap $\Theta\neq1$,
the distinguishability of $\Lambda^{(0)}_{\mathrm{IR}}$ and $\Lambda^{(1)}_{\mathrm{IR}}$ increases in this case, while the error-probability of target detection in the spacetime region decrease.

\begin{figure}[ht]
\centering
\includegraphics[height=2.5in,width=3.2in]{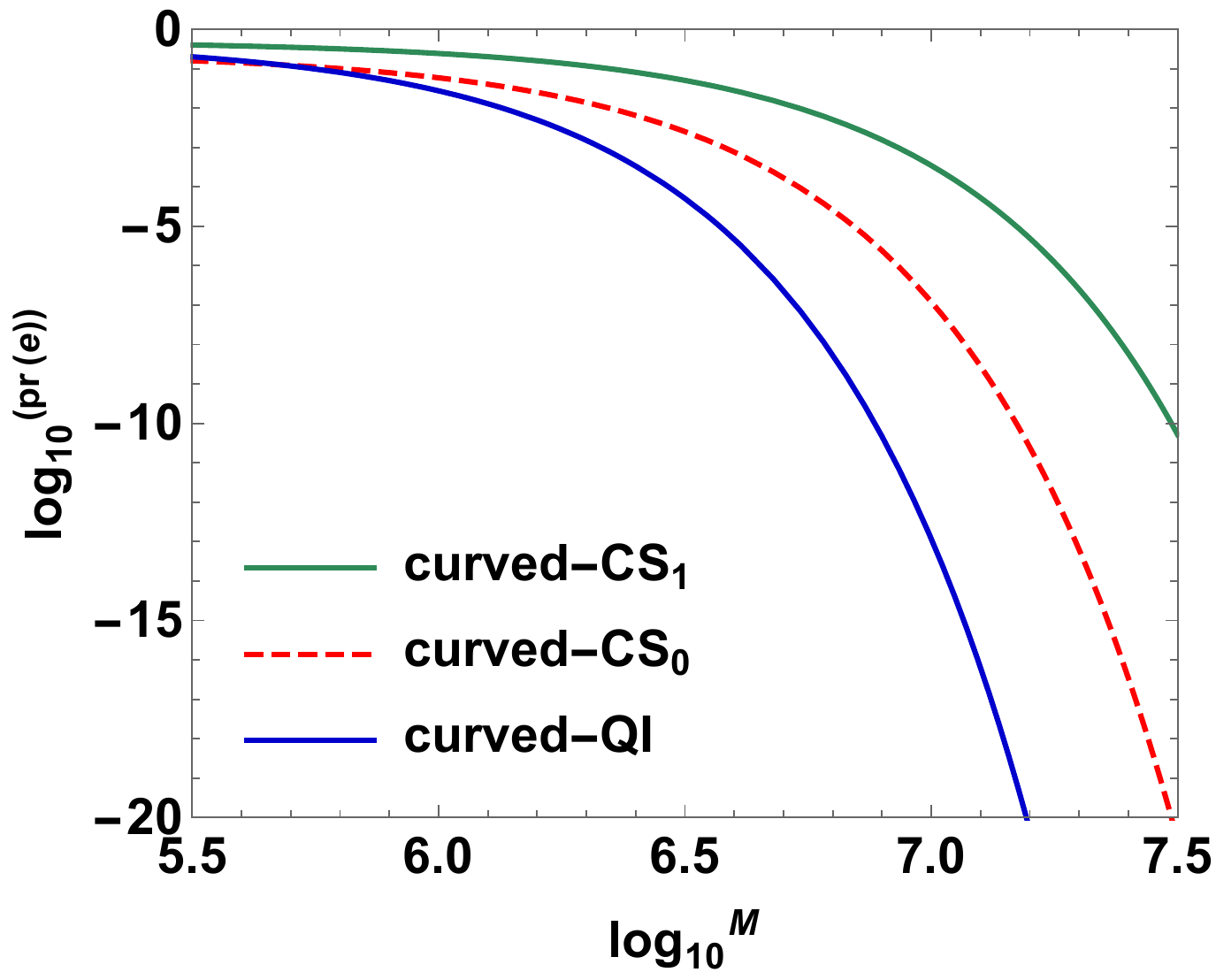}
\caption{(Color online) The bounds on the target detection error-probability as a function of the copies of transmitted modes $M$ (each with $N_S=0.01$ photons on average and $N_B=20$) in the curved spacetime ($R=0.33 \textrm{m}$). }\label{figpare}
\end{figure}

In Fig. (\ref{figpare}) we show the bounds of space target detection strategies  for the fixed range $R=0.33 \textrm{m}$.
When the range is 0.33m, we can experimentally observe the time expansion caused by the Earth's spacetime curvature in the laboratory \cite{atom}.
The green and red dotted curves are  the upper and lower bounds for the coherent-state strategy, and the blue curve is the upper  bound for the QI strategy.
It can be seen that for a given $M$ value, the  error-probability of the entangled state is lower than that of the coherent-state light strategy in near-Earth curved spacetime.
  In other words,  the QI with the entangled state is more efficient than its coherent-state counterpart.
The classical-state system \cite{Tan} with the same energy constraint enables lower error-probability than the coherent-state system, which guarantees a better role
 of the entangled transmitter in spatial target detection. And it behooves us to find a physical explanation for the performance gain provided by QI. In our view, the entangled state with a double-mode structure accounts for this, thus more modes are available for target detection programs. In other words,  more probe states should be used in coherent-state target detection  to improve the distinction between states $\rho^{(0)}$ and $\rho^{(1)}$.

\section{Conclusions} \label{section4}

In this paper we presented the QI transmitter under the background of near-Earth curved spacetime, and investigated the impact of spacetime curvature on the coherent-state strategy and QI strategy.
 The respective effects resulting from the gravitational red-shift and blue-shift on the entangled signal would  cancel each other, while the gravity  always affects  the spatial thermal mode. In this perspective, either for the QI system or the coherent-state system, the detection error-probability in the near-Earth curved spacetime is always lower than that in the flat spacetime case.
 We compared the performance of both kinds of illumination, and it is found that  the spatial quantum target detection with the entangled state is more efficient than its coherent-state counterpart in the near-Earth spacetime.
In conclusion, the spacetime curvature of the Earth  promotes  the detection sensitivity for low-reflectivity target detection by reducing the thermal signal in the returned mode.  This model can be technically applied to microwave frequencies like Ref. \cite{Barzanjeh,Barzanjeh2}.


\begin{acknowledgments}
This work is supported by  the National Natural Science Foundation
of China under Grant  No. 12122504, No.11905218, and No. 11875025;  and the CAS Key Laboratory for Research in Galaxies and Cosmology, Chinese Academy of Science under No. 18010203.

\end{acknowledgments}


\begin{thebibliography}{99}

\bibitem{Lloyd}
S. Lloyd, \textit{Science} {\bf 321}, 1463 (2008).

\bibitem{Tan}
S. H. Tan \emph{et al.}, \textit{Phys. Rev. Lett.} {\bf 101}, 253601 (2008).

\bibitem{Lopaeva}
E. D. Lopaeva, I. Ruo Berchera, I. P. Degiovanni, S. Olivares,
G. Brida, and M. Genovese, \textit{Phys. Rev. Lett.} {\bf 110}, 153603 (2013).

\bibitem{Ragy}
S. Ragy \emph{et al.}, \textit{J. Opt. Soc. Am. B} {\bf 31}, 2045 (2014).





\bibitem{Choi}
J. Choi, V. Va, N. Gonzalez-Prelcic, R. Daniels, C. R. Bhat, and R. W. Heath,  \textit{IEEE Commun. Mag.} {\bf 54}, 160–167 (2016).

\bibitem{Zhang1}
Z. Zhang, M. Tengner, T. Zhong, F. N. C. Wong, and J. H. Shapiro, \textit{Phys. Rev. Lett.} {\bf 111}, 010501 (2013).

\bibitem{Zhang2}
Z. Zhang, S. Mouradian, F. N. C. Wong,  and J. H. Shapiro,  \textit{Phys. Rev. Lett.} {\bf 114}, 110506 (2015).



 \bibitem{Z2}
 Q. Zhuang and S. Pirandola, \textit{Commun. Phys.} {\bf 3}, 103 (2020).

\bibitem{Z3}
 Q. Zhang, \textit{Phys. Rev. Lett.} {\bf 126}, 240501 (2021).

\bibitem{Ren}
L. Maccone and C. Ren, \textit{Phys. Rev. Lett.} {\bf 124}, 200503 (2020).

\bibitem{RK1}
A. Karsa, G. Spedalieri, Q. Zhuang, and S. Pirandola, \textit{Phys. Rev. Res.} {\bf 2}, 023414 (2020).


\bibitem{Barzanjeh}
S. Barzanjeh, S. Guha, C. Weedbrook, D. Vitali, J. H. Shapiro, and  S. Pirandola,
\textit{Phys. Rev. Lett.} {\bf 114}, 080503 (2015).

\bibitem{Barzanjeh2}
 S. Barzanjeh, S. Pirandola, D. Vitali, and J. M. Fink, \textit{ Sci. Adv.} {\bf 6}, eabb0451 (2020).

 \bibitem{earthrev}
R. Howl, L. Hackermuller, D. E. Bruschi, and I. Fuentes, \textit{Adv. Phys.}  {\bf 3}, 1383184 (2018).

\bibitem{earthqs1}
D. E. Bruschi, T. C. Ralph, I. Fuentes, T. Jennewein, and M. Razavi, \textit{Phys. Rev. D}  {\bf 90}, 045041 (2014).



\bibitem{earthqs4}
J. Kohlrus, D. E. Bruschi, J. Louko, and I. Fuentes, \textit{EPJ
Quantum Technol.}  {\bf4}, 7 (2017).

\bibitem{earthqs2}
D. E. Bruschi, A. Datta, R. Ursin, T. C. Ralph, and I. Fuentes, \textit{Phys. Rev. D}  {\bf 90}, 124001 (2014).


\bibitem{earthqs5}
J. Kohlrus, D. E. Bruschi, and I. Fuentes, \textit{Phys. Rev. A}  {\bf 99}, 032350 (2019).

\bibitem{earthqs6}
S. P. Kish and T. C. Ralph, \textit{Phys. Rev. D}  {\bf 99}, 124015 (2019).

\bibitem{earthqs3}
T. Liu, J. Jing, and J. Wang,  \textit{Adv. Quantum Technol.}  {\bf1}, 1800072 (2018).

\bibitem{earthqs7}
T. Liu, S. Cao, and S. Wu, \textit{Sci. Rep.}  {\bf10}, 14697 (2020).

\bibitem{QC}
Q. Liu, C. Wen, J.  Wang, and J.  Jing, \textit{Ann.Phys.}  {\bf 533}, 2000536 (2021).

\bibitem{wangqcs}
J. Wang, Z. Tian, J.  Jing, and Heng Fan, \textit{Phys. Rev. D}  {\bf 93}, 065008 (2016).



\bibitem{Rideout}
D. Rideout \emph{et al.}, \textit{Classical Quantum Gravity} {\bf 29}, 224011 (2012).

\bibitem{Zych}
M. Zych, F. Costa, I. Pikovski, and C. Brukner, \textit{Nat. Commun.} {\bf 2}, 505 (2011).


\bibitem{Visser}
M. Visser, 2007, arXiv:0706.0622.

\bibitem{Downes}
T. G. Downes, T. C. Ralph, and N. Walk, \textit{Phys. Rev. A} {\bf 87}, 012327 (2013).

\bibitem{Leonhardt}
U. Leonhardt,  \textit{Measurement Science and Technology} {\bf 11}, 1827 (2000).

\bibitem{DE}
 D. E. Bruschi, S. Chatzinotas, F. K. Wilhelm, and A. W. Schell,  \textit{ Phys. Rev. D } {\bf 104}, 085015 (2021).

\bibitem{DE2}
 D. E. Bruschi and A. W. Schell,  \textit{ arXiv:2109.00728 }.

\bibitem{Rohde}
P. P. Rohde, W. Mauerer, and C. Silberhorn, \textit{New J. Phys.} {\bf 9}, 91 (2007).

\bibitem{700THZ}
D. N. Matsukevich, P . Maunz, D. L. Moehring, S. Olmschenk, and C. Monroe, \textit{Phys. Rev. Lett.} {\bf 100},
150404  (2008).

\bibitem{story}
J. H. Shapiro,  \textit{IEE Aerospace and EI ectronic Systems Magazine} {\bf 35}, 8-20 (2020).

\bibitem{Wong}
F. N. C. Wong, J. H. Shapiro, and T. Kim, \textit{Laser Phys.} {\bf 16}, 1517 (2006).



\bibitem{Helstrom}
C. W. Helstrom, \textit{Inf. Control} {\bf 10}, 254 (1967).

\bibitem{Audenaert}
K. M. R. Audenaert \emph{et al.},
\textit{Phys. Rev. Lett.} {\bf 98}, 160501 (2007).

\bibitem{Nussbaum}
K. M. R. Audenaert, M. Nussbaum, A. Szkola, and F. Verstraete, \textit{Commun. Math. Phys.} {\bf 279}, 251 (2008).

\bibitem{Calsamiglia}
J. Calsamiglia, R. Munoz-Tapia, L. Masanes, A. Acin, and E. Bagan, \textit{Phys. Rev. A } {\bf 77}, 032311 (2008).

\bibitem{Kailath}
T. Kailath, \textit{IEEE Trans. Commun. Technol.} {\bf 15}, 52 (1967).

\bibitem{Pirandola}
S. Pirandola and S. Lloyd, \textit{Phys. Rev. A} {\bf 78}, 012331 (2008).


\bibitem{Weedbrook}
C. Weedbrook, S. Pirandola, R. Garcia-Patr\'{o}n, N. J. Cerf, T. C. Ralph, J. H. Shapiro, and S. Lloyd,
\textit{Rev. Mod. Phys.} {\bf 84}, 621 (2012).

\bibitem{atom}
C. W. Chou, D. B. Hume, T. Rosenband, and D. J. Wineland,  \textit{Science} {\bf 329}, 1630 (2010).

\bibitem{Ivette}
I. Fuentes-Schuller and R. B. Mann, \textit{Phys. Rev. Lett.} {\bf 95}, 120404  (2005).


\bibitem{Wang2}
J. Wang, H. Cao, J. Jing, and H. Fan, \textit{Phys. Rev. D} {\bf 93}, 125011 (2016).

\bibitem{David3}
D. E. Bruschi, I. Fuentes, and J. Louko, 	\textit{Phys. Rev. D} {\bf 85}, 061701(R) (2012).

\bibitem{Tian}
Z. Tian, J. Wang, J. Jing, and A. Dragan, \textit{ Ann. Phys.} {\bf 377}, 1-9 (2017).

\bibitem{JW}
 J. Wang, C. Wen, S. Chen, and J. Jing,  \textit{Phys. Lett. B} {\bf 800}, 135109 (2020).

\bibitem{Hawking}
S. W. Hawking, \textit{Nature} {\bf 248}, 30 (1974); S. W. Hawking, \textit{Commun. Math. Phys.} {\bf 43}, 199 (1975).


\end{thebibliography}
\end{document}